\documentclass[aps,twocolumn,superscriptaddress,showpacs]{revtex4}
\usepackage{graphics,graphicx,dcolumn,bm,fleqn,epic,eepic,float}
\usepackage{amssymb,amsmath,multirow,rotate,color,float,times}

\definecolor{blue}{rgb}{0,0,1}
\definecolor{red}{rgb}{1,0,0}

\begin{document}
\title{Obtaining the 
       size distribution of fault gouges with polydisperse 
       bearings}

\date{\today}

\author{Pedro G.~Lind}
\affiliation{Institute for Computational Physics, 
             Universit\"at Stuttgart, Pfaffenwaldring 27, 
             D-70569 Stuttgart, Germany}
\author{Reza M.~Baram}
\affiliation{Computational Physics,
             IfB, HIF E12, ETH H\"onggerberg, CH-8093 Z\"urich, Switzerland}
\author{Hans J.~Herrmann}
\affiliation{Computational Physics,
             IfB, HIF E12, ETH H\"onggerberg, CH-8093 Z\"urich, Switzerland}
\affiliation{Departamento de F\'{\i}sica, Universidade Federal do Cear\'a,
             60451-970 Fortaleza, Brazil}

\begin{abstract}
We generalize the recent study of random space-filling bearings to
a more realistic situation, where the spacing offset varies
randomly during the space-filling procedure, and 
show that it reproduces well the size-distributions 
observed in recent studies of real fault gouges.
In particular, we show that the fractal dimensions of random
polydisperse bearings sweep predominantly the low range of values 
in the spectrum of fractal dimensions observed along real faults,
which strengthen the evidence that polydisperse bearings may explain
the occurrence of seismic gaps in nature.
In addition, the influence of different distributions for the 
offset is studied and we find that the uniform distribution is the 
best choice for reproducing the size-distribution of fault gouges.
\end{abstract}


\maketitle


\section{Introduction}

\begin{figure}[htb]
\begin{center}
\includegraphics*[width=5.0cm]{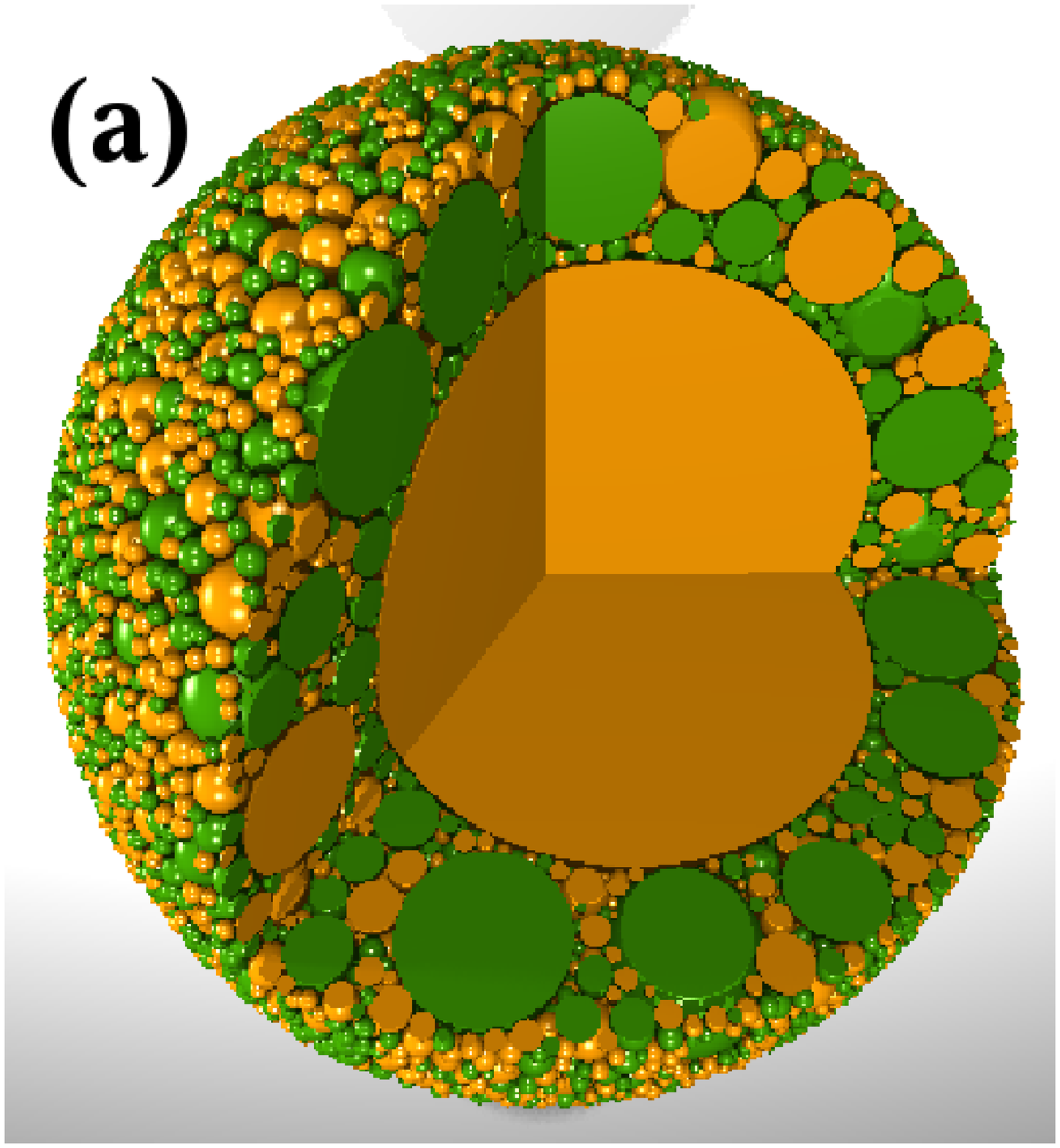}
\includegraphics*[width=5.0cm]{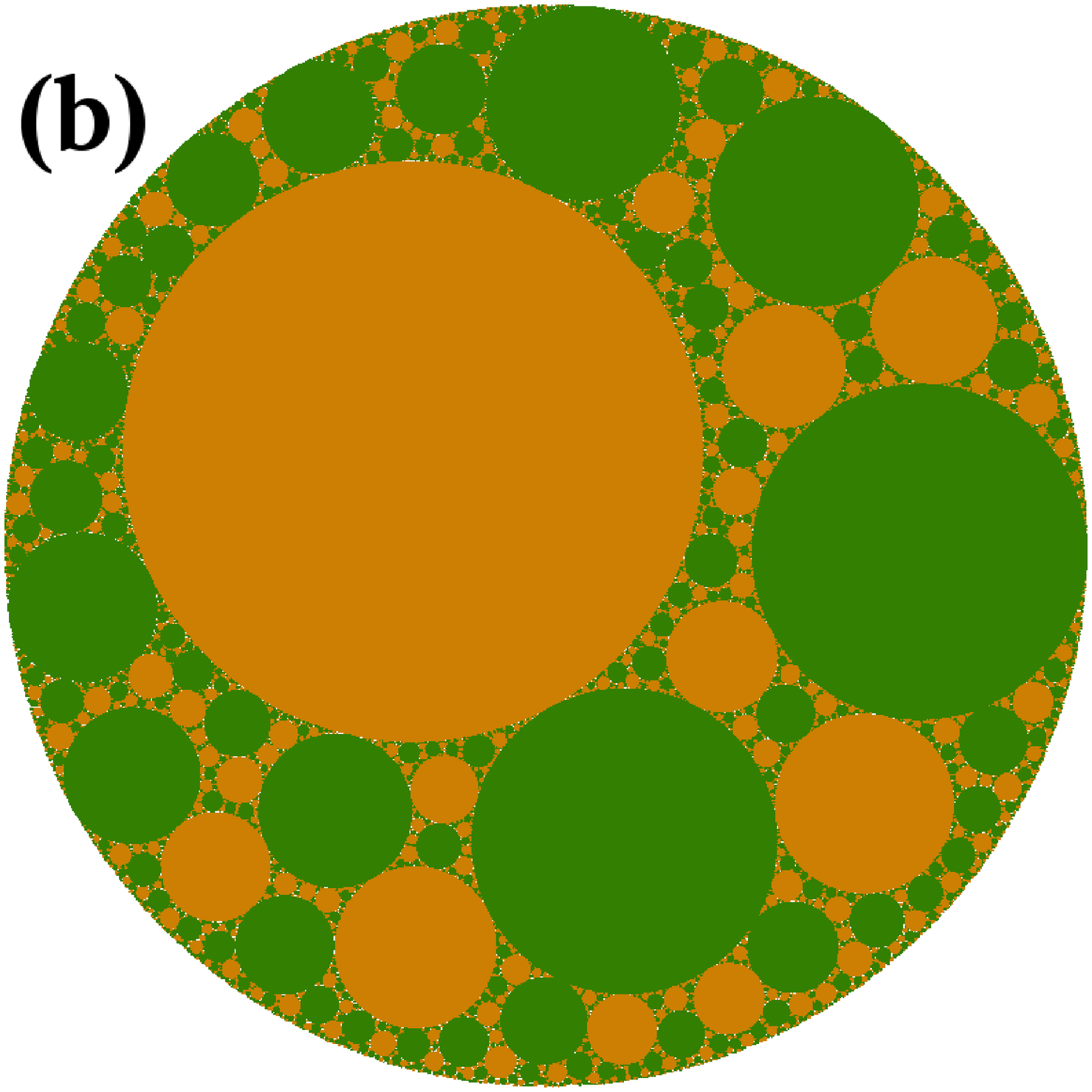}
\end{center}
\caption{\protect 
  (Color Online)
  Illustration of a random space-filling bearing in
  {\bf (a)} three and in
  {\bf (b)} two dimensions.
  Discs and spheres of the same color do not touch
  each other.
  The random space-filling bearing starts with a large
  disc or sphere, maximizing polydispersity (see text).}
\label{fig1}
\end{figure}

In the early nineties the question of the possibility to tile
an arbitrarily large strip of space-filling roller bearings
without friction nor slipping was addressed~\cite{herrmann90}, 
motivated by the study of real systems such as seismic gaps.
Seismic gaps are regions along a fault zone where earthquakes 
do not take place and therefore they could be explained by
sheared plates on a space-filling bearing~\cite{astrom00}.
Fault zones are typically self-similar and the mechanical origin 
of the power-law in the particle size distribution was 
associated to the particle's fracture probability which has been 
proposed to be controlled by the relative size of its nearest 
neighbors~\cite{sammis87,sammis89}.
More recently, a geophysical model~\cite{sammis07} explained the 
different values of the fractal dimension,
ranging from $d_f=2.6$ to $d_f\simeq 3$, by taking
into account the fault gouge strain.
While such model explains the dynamical origin of different 
power-laws, there is still the question if the
space-filling bearing scenario is able to reproduce
such empirical results in a simple and systematic way.
By reproducing with space-filling bearings the same particle 
size distributions observed in fault zones one can strengthen
the hypothesis that the existence of seismic gaps in fault
zones may be related to the emergence of particular geometrical 
arrangements of their composing rocks, due to local fragmentation
during tectonic motion.
Figure \ref{fig1} illustrates two random space-filling bearings in
two and three dimensions.

Pioneering studies with space-filling bearings were done using 
deterministic procedures in two~\cite{herrmann90} and 
three~\cite{baram04b} dimensions and also using random 
algorithms~\cite{baram05}.
However, up to now specific initial configurations and fixed
parameter values were addressed.
In this paper we present a general algorithm to construct 
realistic space-filling bearings that allows to reproduce the 
range of fractal dimensions 
observed in fault gouges~\cite{sammis87,sammis89,sammis07,roux93}.
The model is parameterized by a unique parameter that controls
the strength of fragmentation and takes into account ensemble
averages. 
Despite the wider freedom in the parameters
and initial configurations, the systems presents robust results in 
what concerns the fractal dimension.
In particular, we will show that by varying the range of admissible
values of the control parameter one finds fractal dimensions observed 
in fault gouges.

We start in Section \ref{sec:prevalg} by describing in some detail
the procedure to generate random space-filling bearings, introducing
a parameter that accounts for fragmentation at the local scale.
In Sec.~\ref{sec:2D} and \ref{sec:3D} we describe the results
for two and three dimensions respectively, with special emphasis 
on the size distribution and the fractal dimension.
Discussions and conclusions are given in Sec.~\ref{sec:conc}.

\section{The random space-filling of particles}
\label{sec:prevalg}

In this Section we will start by revisiting previous
procedures~\cite{baram05} for constructing random space-filling 
packings and bearings and then introduce the necessary 
ingredients to obtain a fully random space-filling bearing.

\begin{figure}[b]
\begin{center}
\includegraphics*[width=4.0cm]{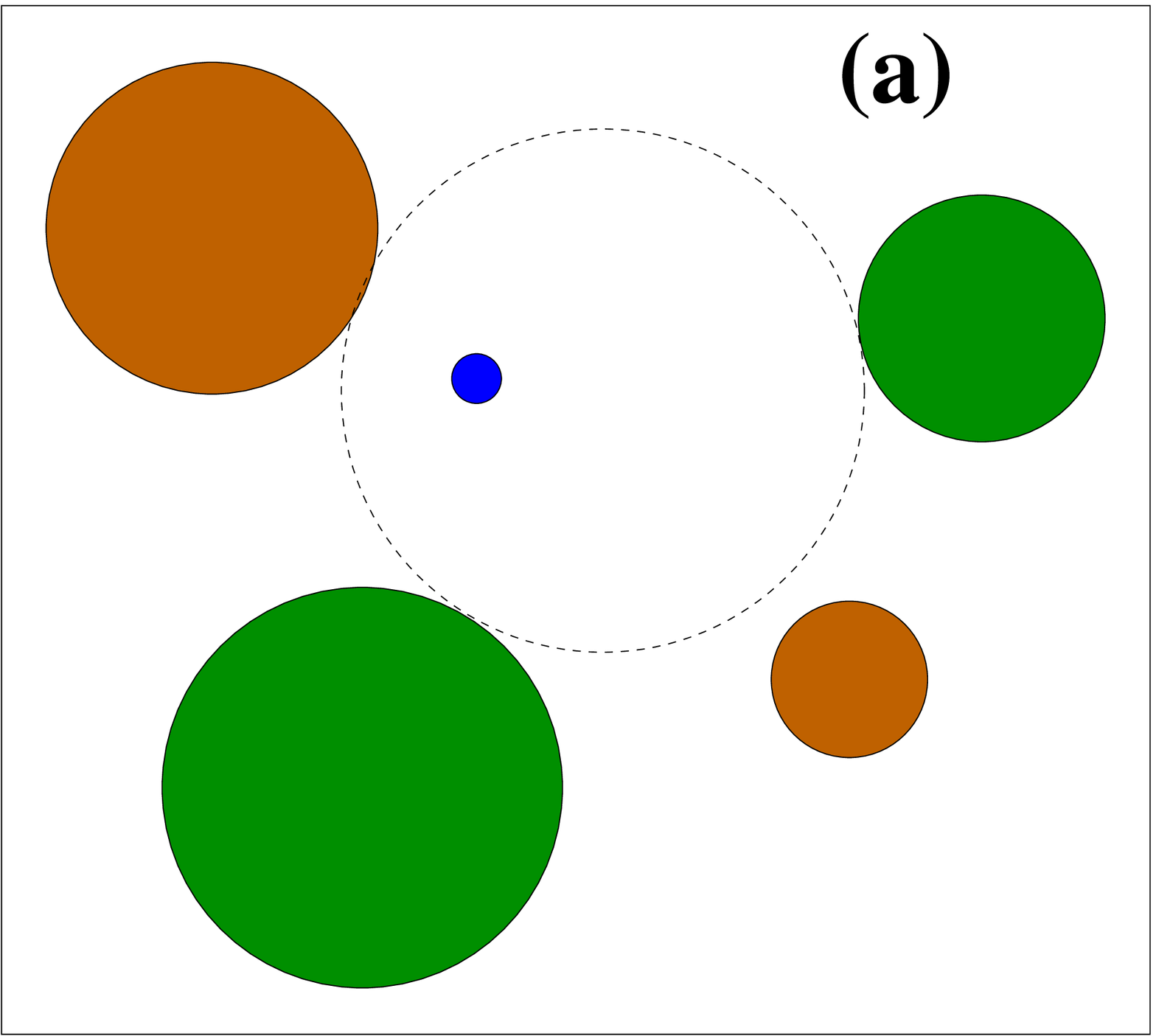}%
\includegraphics*[width=4.0cm]{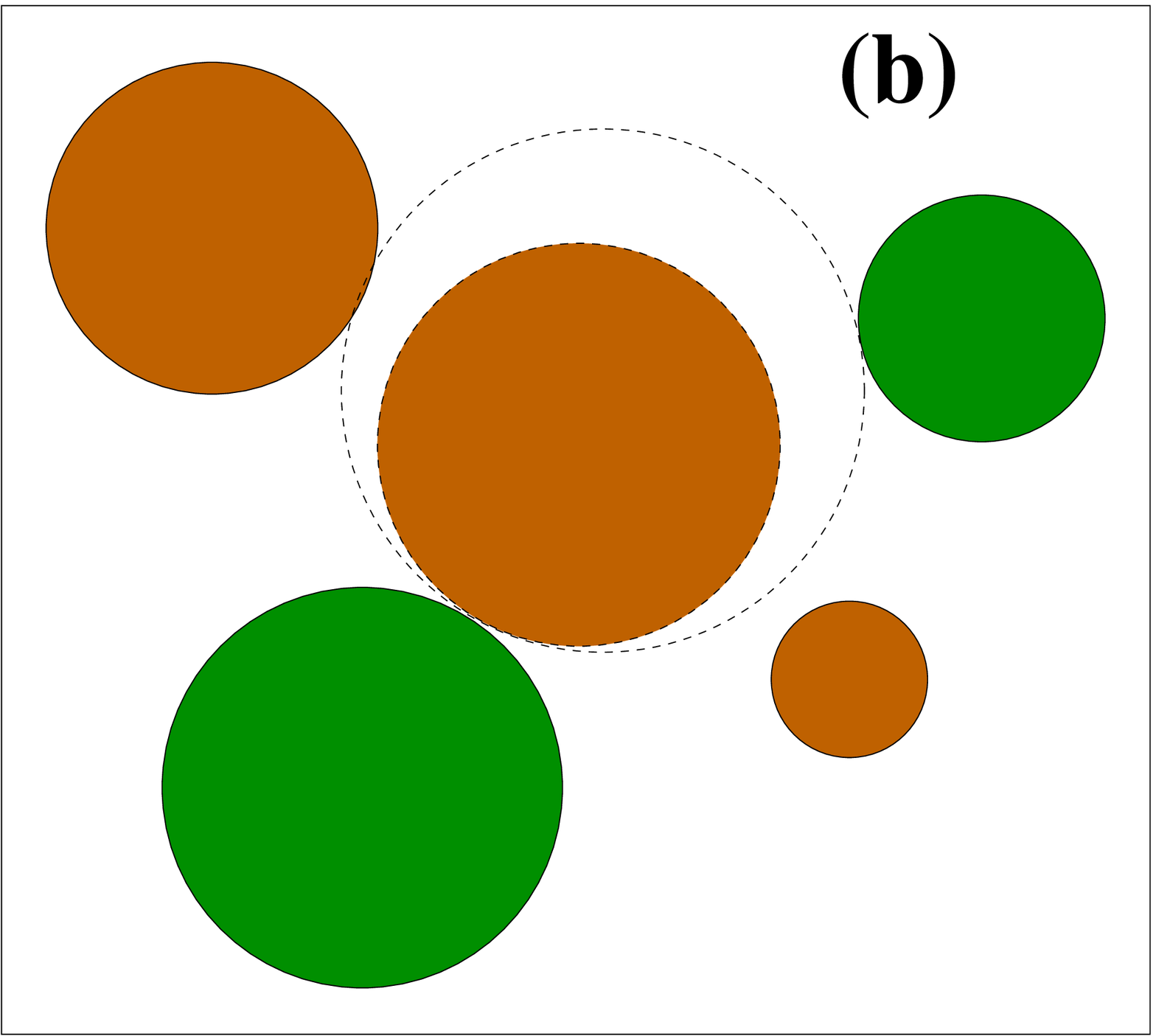}
\includegraphics*[width=4.0cm]{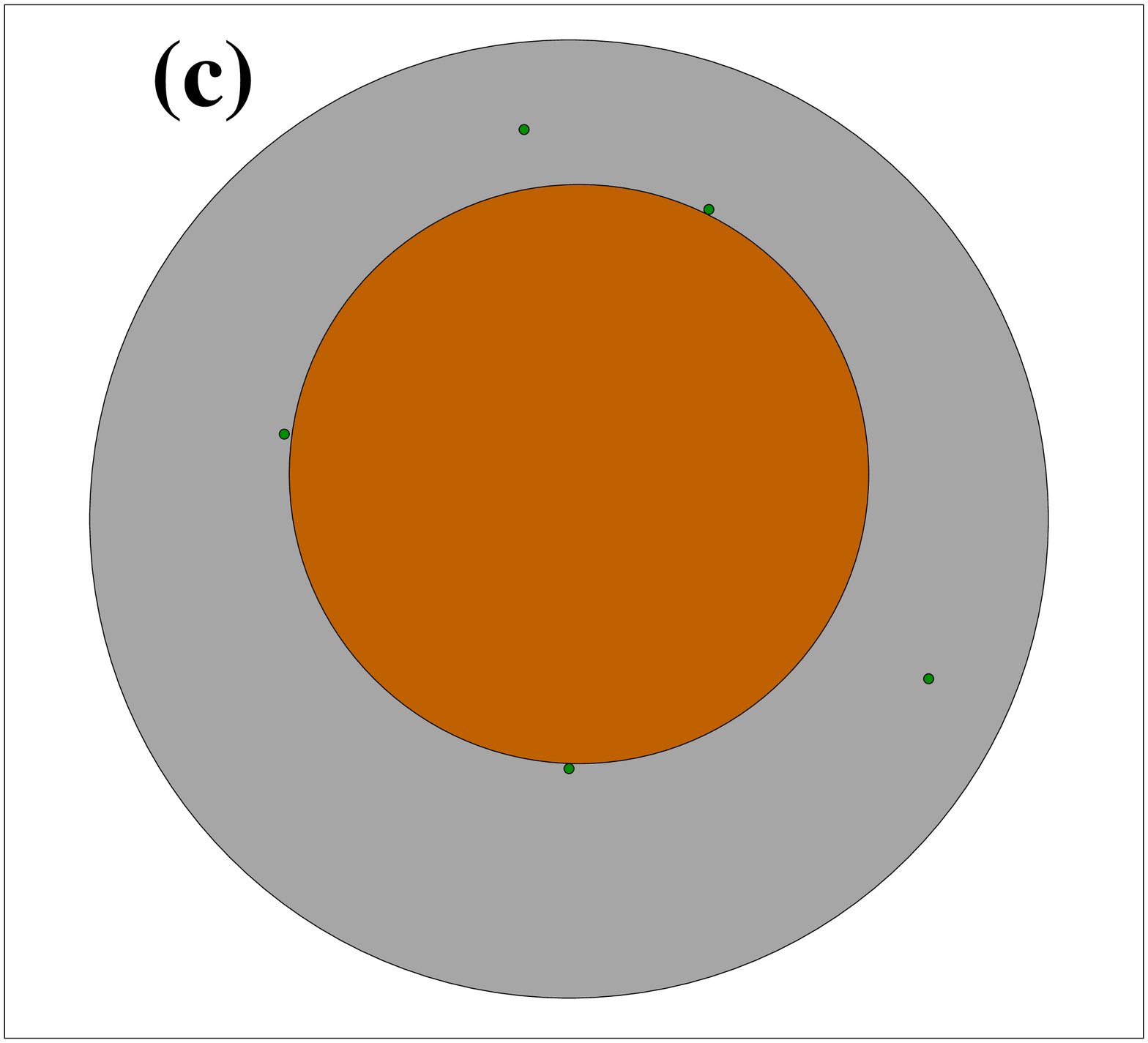}
\end{center}
\caption{\protect 
  (Color Online)
  Sketch of the construction of a random space-filling
  bearing.
  {\bf (a)} A new disc (blue) is randomly inserted in the system
  and is shifted and enlarged to the maximal accessible size 
  (dashed circle) without overlapping neighboring discs.
  Then,
  {\bf (b)} it is reduced by a factor $0<\alpha\le 1$, keeping a 
  single contact point with one of the neighbors and assuming the 
  opposite color.
  {\bf (c)} To start the space-filling with a large disc (brown)
  one needs to place previously a few small `seeds' (green) in the 
  system (grey) and then proceed as in (a) and (b) [see text].}
\label{fig2}
\end{figure}

Random bearings in two and three dimensions are constructed in the 
following way.
First, one starts by randomly distributing a small number 
$N_0$ of discs or spheres within a given range of sizes, without 
touching each other.
Second, one fills the empty spaces in the system by introducing
iteratively the biggest possible disc or sphere in the neighborhood of
some empty region.
Third, one resizes some disc or sphere in order for the packing 
to be bi-chromatic (bearing condition), i.e.~only two colors are needed 
to color all discs in such a way that no discs of the same color touch 
each other. This guarantees the bearing condition: particles are able 
to role on each other without friction or slipping. 
Figures \ref{fig2}a and \ref{fig2}b give illustrative examples of such
random bearings in two dimensions.

The filling procedure is done by choosing randomly a void 
within the inter-disc free space and then fitting the biggest disc 
in it, i.e.~fit the disc that touches the three nearest discs 
in the neighborhood, as illustrated in Fig.~\ref{fig2}a.
For the three dimensional case one considers spheres touching the
four nearest neighbors.

The coloring procedure is done by attributing a proper color to the 
introduced disc.
In the case that the three neighboring discs have the same
color one attributes the other color to the new disc.
Otherwise, one chooses only one of the neighbors to be in contact
with the new disc, and the new disc shrinks to a size with
radius $r=\alpha r_0$ ($0\le \alpha\le 1$), where $r_0$ is the radius 
before shrinking and gets a different color as the disk it touches.
Figure \ref{fig2}b illustrates this coloring procedure.

Parameter $\alpha$ is our control parameter.
For constant $\alpha=1$ one obtains the particular case where bearing 
cannot be guaranted. In this case, frustrated contacts emerge when 
particles are forced to rotate~\cite{baram05}, which would eventually 
lead to the fragmentation of the discs into smaller ones. 
Recently a new method to implement realistic grain fracture in
three-dimensional simulations of granular shear was proposed~\cite{abe05},
based on breakable bonds between particles within a medium.
We keep the model simple, by using instead the reduction
factor $\alpha$ that mimics the effect of fragmentation: by shrinking
a particle originally with frustrated contacts we mimic its fragmentation
into smaller particles that will fill the empty space left after the
fragmentation.

The algorithm described above was previously\cite{baram05,baram04c} 
used in two and three dimensions by fixing a given initial 
configuration with $N_0$ discs having sizes within a given fixed range 
and also by using a fixed reduction factor $\alpha$ during the filling
and coloring stages.
The density of the packing was studied as a function of the number $N$ 
of existing spheres as well as the cumulative particle size distribution.
It was found that the cumulative distribution obeys a power-law, 
namely $N(r)\equiv \int_r^{\infty} n(q)dq \sim r^{-d_f}$, where
$d_f$ is the fractal dimension of the bearing~\cite{manna91}.

Next, we introduce the additional points to strengthen previous findings 
and improve the algorithm described above.

First, there is the statistical significance of the results, 
and their sensitiveness to initial configurations, i.e.~to the
initial range $[r^{\ast}-\delta/2,r^{\ast}+\delta/2]$ of sizes.
This initial configuration may influence the polydispersity of the system
and consequently the attained distribution after filling the entire 
system.
As we will see, large number $N_0$ of initial discs typically
influence how the density increases during the space-filling procedure.
\begin{figure*}[t]
\begin{center}
\includegraphics*[width=15.0cm]{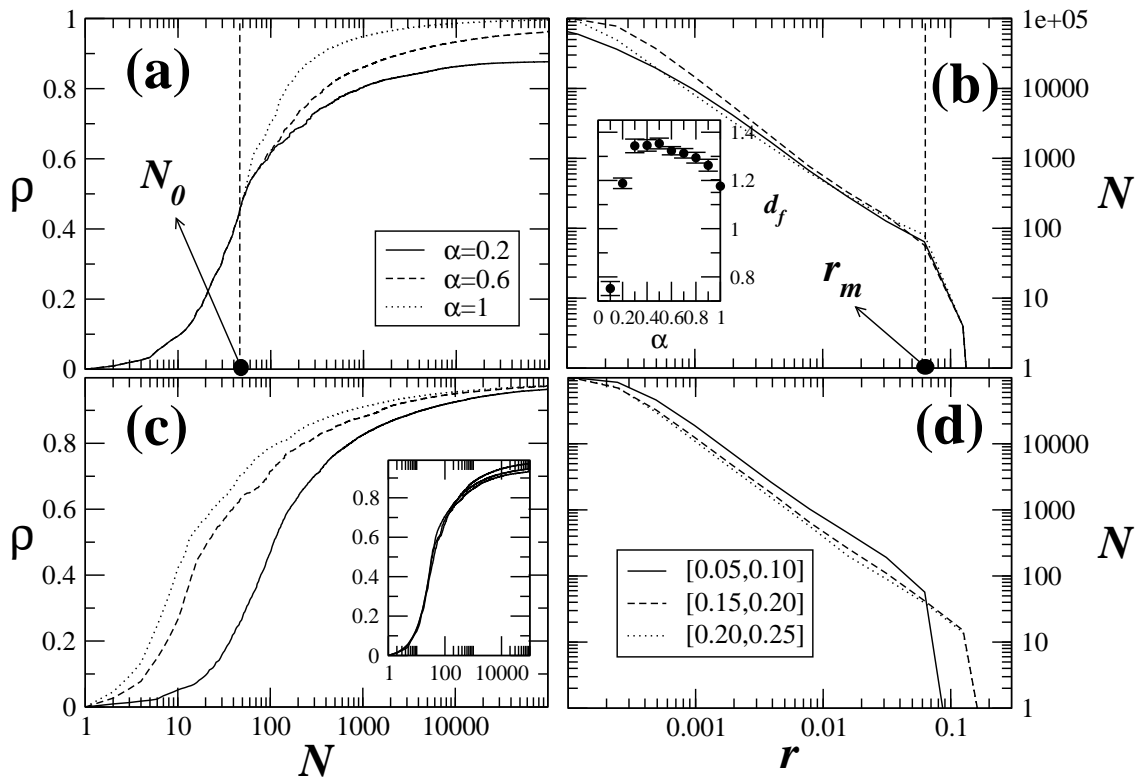}
\end{center}
\caption{\protect 
  Density and size distribution in two-dimensional random space-filling 
  packings.
  {\bf (a)} Density $\rho$ of the packing as a function of the total 
  number of discs $N$ for different values of $\alpha=0.2,0.6$ and $1.0$, 
  together with
  {\bf (b)} the distribution of the radius $r$ of the discs.
  In both cases, one starts from a fixed initial configuration of
  $N_0=40$ discs with radius in the range $[0.08R,0.14R]$ with $R=1$ 
  being the radius of the system (see Fig.~\ref{fig1}b).  
  The minimal radius of the initial set of discs is indicated as
  $r_m\sim 0.08$ (see text). 
  Fitting of the power-law range in (b), the distribution $N(r)=br^{-d_f}$ 
  yields the fractal
  dimension $d_f$ as a function of $\alpha$ plotted in the inset.
  In {\bf (c)} one plots the density as a function of $N$ for
  the initial sets of $N_0=40$ discs in $[0.05R,0.10R], 
  [0.15R,0.20R]$ and $[0.20R,0.25R]$, i.e.~ranges with the same
  interval width but centered around different values, namely around
  $r^{\ast}=0.075, 0.175$ and $0.225$ respectively and fixing
  $\alpha=0.6$.
  In the inset of (c) the density $\rho(N)$ is plotted for
  initial ranges $[r^{\ast}-\tfrac{\delta}{2},r^{\ast}+\tfrac{\delta}{2}]$
  having the same center $r^{\ast}=0.075$ but different widths 
  $\delta=0.2, 0.4, 0.6$ and $0.8$. 
  In {\bf (d)} we plot distribution $N(r)$ of the spheres as a function 
  of $r$ for the same conditions as in (c). In all cases $N=10^5$ discs.}
\label{fig3}
\end{figure*}

To take this point into account we enable the construction
procedure to start with an initial configuration having a single 
arbitrarily large disc (or sphere) and perform ensemble averages 
on a significant number of initial configurations.
Concerning the single initial large disc, one should notice that 
it does not suffice one single disc to introduce a second one, 
because in two-dimensions each inserted disc needs to have at least
three neighbors (four neighbours for three-dimensions).
However, as illustrated in Fig.~\ref{fig2}c, such starting disc 
can be introduced
into the system by previously distributing a few very small `seed-discs'
in the system and then following the algorithm described above.
The number of such seeds is small, and therefore they do not affect
significantly the cumulative size distribution.
Their role is that the average distance between them 
is eventually of the order of the system size, enabling the
introduction of a first disc with the size of the order of the
system size.
Of course, depending on the number and distribution of the seeds,
the first initial discs may have also a size within the 
initial range of sizes.
In this way, one generalizes the previous procedure~\cite{baram05} and 
maximizes the admissible polydispersity.

Second, we also introduced a criterion to increase computational 
efficiency of our algorithm. The neighborhood were the neighboring
discs (or spheres) are searched for must be chosen conveniently. 
We propose to choose a size that decreases with the increase
of the density $\rho$, since the denser the packing the smaller are 
the empty spaces to put new discs. Therefore, the radius $r_n$ of the 
neighborhood of a given random point introduced in the system
at iteration $n$, is updated as $r_n = \frac{1-\rho_n}{\rho_n} 
(r_{sys}-r_{max})$, where $r_{sys}$ and $r_{max}$ are the radii
of the system and of the biggest disc or sphere in it, respectively.

Third, we also consider the control parameter $\alpha$ to vary
randomly within a tunable range of values.
In particular, we argue that though a tentative value of constant 
$\alpha$ could be obtained by analyzing samples of gouges in real 
situations, one expects that a certain range of admissible values 
for $\alpha$ is the most realistic assumption. Indeed, we show
that the typical range of fractal dimensions observed in real fault 
gouges is in this way reproduced.

\section{The two-dimensional case}
\label{sec:2D}

We start this section by addressing the two-dimensional random
space-filling bearing and systematically reviewing the behavior of 
the packing for different, but fixed, values of $\alpha$ and study
the effect of fixed initial size ranges (no maximal admissible
polydispersity). 
Figure \ref{fig3} shows for this case the density $\rho(N)$ and 
cumulative size distribution $N(r)$ of two-dimensional random 
space-filling packings. 
\begin{figure*}[htb]
\begin{center}
\includegraphics[width=15.0cm]{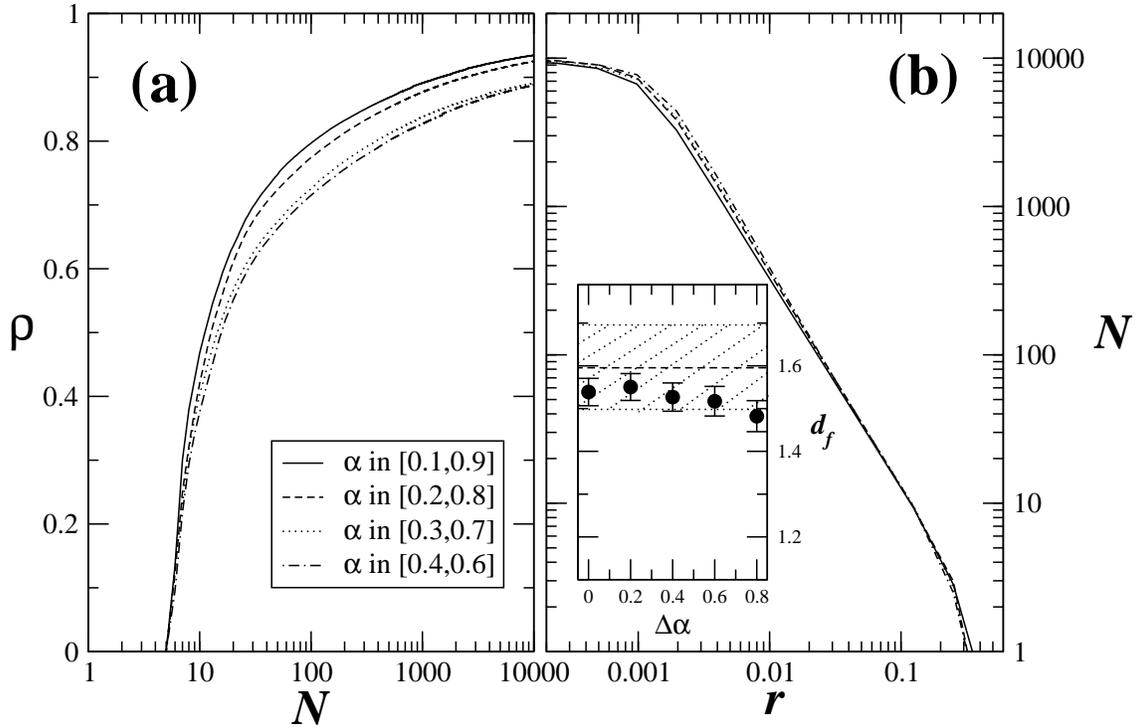}
\end{center}
\caption{\protect 
  Averaging 
  {\bf (a)} the density $\rho$ and 
  {\bf (b)} the distribution size $N(r)$ over $100$
  initial configurations starting from large discs 
  ($r\sim R/2$) and with $\alpha$ varying in
  a range $[0.5-\Delta\alpha/2,0.5+\Delta\alpha/2]$ with
  $\Delta\alpha=0.2,0.4,0.6$ and $0.8$. The inset of (b)
  shows that the fractal dimension in 
  $N(r)\sim r^{-d_f}$ is almost
  independent of $\Delta\alpha$ yielding $d_f\simeq 1.54$ which is,
  within the numerical errors of real fault gouges (see text).}
\label{fig4}
\end{figure*}

Figure \ref{fig3}a shows the density $\rho$ as a function of the number 
$N$ of discs for $\alpha=1.0$ (packing) and also for $\alpha=0.2$ and 
$0.6$ (bearings) separately, starting from $N_0=40$ discs with radius 
in the range $[r^{\ast}-\tfrac{\delta}{2},r^{\ast}+\tfrac{\delta}{2}]$
having $r^{\ast}=0.11$ and $\delta=0.06$.
As expected, the convergence $\rho\to 1$ as $N$ increases is
faster for larger values of $\alpha$.
We consider one fixed initial configuration with $N_0$ initial
discs, and therefore the different curves coincide for $N<N_0$. 

In Fig.~\ref{fig3}b we plot the distribution of the radius $r$ of the 
discs, where $r_m=r^{\ast}-\tfrac{\delta}{2}$ is the minimal radius of 
the initial set of discs, and the deviation from the power law for $r>r_m$ 
is due to the initial configuration.
Below this value $r_m$, the size distribution obeys a power-law
$N=br^{-d_f}$, where $d_f$ is the fractal dimension of the packing,
plotted in the inset as a function of  $\alpha$ (symbols).
As one sees from the inset, the fractal dimension typically takes values 
in the range $1.2 < d_f < 1.4$, differently from the values found in
two-dimensional cuts of fault gouges ($\sim 1.6\pm 0.1$).
There is a maximum of $d_f$ for $\alpha=0.5$ that can be explained
from the definition of $\alpha$ in the algorithm described above.
For an $\alpha<0.5$, to each new disc introduced there is a remaining free 
space characterized by $\alpha^{\prime}>0.5$ such that 
$\alpha+\alpha^{\prime}=1$, and similarly for $\alpha>0.5$. 

Both Figs.~\ref{fig3}a and \ref{fig3}b consider the same initial configuration.
To study the influence of the initial configurations, we plot in 
Fig.~\ref{fig3}c the density $\rho(N)$, fixing
$\alpha=0.6$, similar to previous works\cite{baram04c}, and using different 
size ranges for the initial sets of $N_0=40$ discs, namely in $[0.05R,0.10R], 
[0.15R,0.20R]$ and $[0.20R,0.25R]$, i.e.~ranges with the same
width $\delta=0.06$ but centered around different values, namely around
$r^{\ast}=0.075, 0.175$ and $0.225$ respectively.

Since different initial configurations are now used, the density is
no longer the same below $N_0$ as in Fig.~\ref{fig3}a.
Further, one observes that the density converges to one for increasing
the value of $r^{\ast}$.
In the inset the density $\rho(N)$ is plotted by fixing $r^{\ast}=0.075$ 
and starting with initial configuration having different widths, namely 
$\delta=0.2, 0.4, 0.6$ and $0.8$. 
In these cases the density gives always similar dependencies on $N$. 
Therefore, the average size $r^{\ast}$ of the initial configuration
is the important parameter to tune the density of the packing. Its width 
can be varied without changing significantly the results.

In Fig.~\ref{fig3}d we plot the accumulative size distribution 
$N(r)$ of the discs for the same conditions as in Fig.~\ref{fig3}c.
The value of the exponent remains almost constant, $d_f \sim 1.35$.
In other words, the fractal dimension is not very sensitive to the initial
configuration, and the parameter on which the fractal dimension depends
more strongly must be indeed $\alpha$.
\begin{figure}[htb]
\begin{center}
\includegraphics[width=8.5cm]{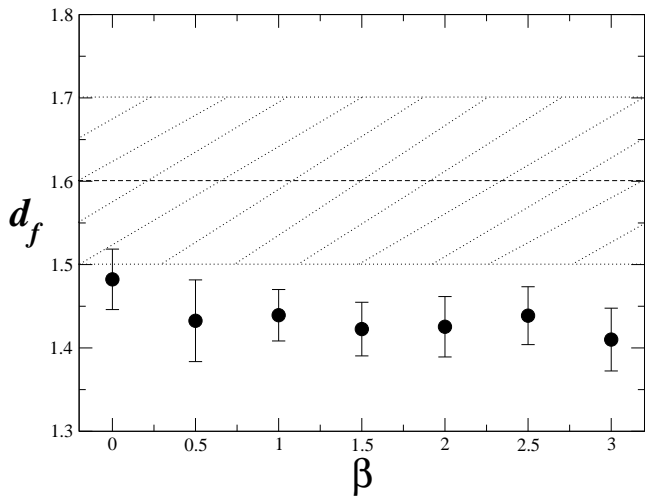}
\end{center}
\caption{\protect 
   The fractal dimension as a function of the exponent
   $\beta$ when the value of $\alpha$ is chosen according to
   a power-law $P(\alpha)\sim \alpha^{-\beta}$ in a range 
   $\alpha\in[0.1,0.9]$. 
   Although within the error bars, the fractal dimension is 
   somewhat lower compared to Fig.~\ref{fig4}b where the 
   distribution of chosen $\alpha$ values is uniform 
   (see text).}
\label{fig5}
\end{figure}

Since $\alpha$ is also the parameter controlling the fragmentation
of discs with frustrated contacts (see above), we will now study it 
more deeply.
When $\alpha$ is able to vary randomly, the fragmentation of the largest disc 
in the free holes can be regarded as a random process by its own. 
We next consider $\alpha$ to be each time randomly selected from a fixed 
interval $[\alpha^{\ast}-\Delta\alpha/2,\alpha^{\ast}+\Delta\alpha/2]$.
We will show that when enabling $\alpha$ to take different values
for each particle shrinking, one obtains fractal dimensions similar 
to the ones observed in fault gouges\cite{sammis87,sammis89,sammis07}.

To this end, we put everything together, namely $\alpha$ varying in the
middle range of admissible values, a large initial disc 
and an ensemble average over a significant number of initial 
configurations.
The results for density and size distribution are shown in Fig.~\ref{fig4}, 
where one considers three initial seeds in the range $r_M=2r_m\sim R/1000$, 
with $R$ the size of the system and $\alpha$ varies randomly in the
range $[0.5-\Delta\alpha/2,0.5+\Delta\alpha/2]$. Averages over a sample 
of $100$ initial configurations.
Since initialization, filling and coloring procedure are now all
random, we call these systems fully random space-filling bearings.

Figure \ref{fig4}a shows the density as a function of the number $N$ 
of discs for $\Delta\alpha=0.2,0.4,0.6$ and $0.8$. One
sees an abrupt transition above $N=3$ (initial seeds), due to the introduction
of the first large disc.
For all the four cases the dependence of $\rho$ on
the range of $\alpha$-values is similar, with the convergence 
towards $\rho=1$ being slightly slower for narrower ranges, because they 
hinder the occurrence of large discs. 

Figure \ref{fig4}b shows the size distribution $N(r)$ for each 
of the four ranges. All the distributions almost coincide, as shown
in the inset where the fractal dimension $d_f$ taken from $N(r)\sim r^{-d_f}$ 
is almost constant ($d_f\sim 1.54$).
This value is larger than the one obtained when $\alpha$ is kept constant 
(see Fig.~\ref{fig3}).
Notice that the value of the fractal dimension for $\Delta\alpha=0$, though
corresponding to the case of constant $\alpha=0.5$, is different from the 
one plotted in the inset of Fig.~\ref{fig3}b, since the constructing
procedure of the bearing is slightly different (see Sec.~\ref{sec:prevalg}).

Since the above value is obtained from a significantly larger sample
of initial configurations and all the parameters $\alpha$ and position
of the discs are randomly selected, we will consider this value
$\bar{d}_f=1.54$ as the characteristic exponent of the size
distribution for fully random two-dimensional space filling bearings.
The average characteristic value $\bar{d}_f=1.54$ obtained lies in the 
range of values measured of the fractal dimension measured in real fault 
gouges ($D=1.6\pm 0.1$)~\cite{sammis87,sammis89}, as indicated with
a dashed line and shadow region in the inset of Fig.~\ref{fig4}b.

Till now, the value of $\alpha$ was considered to vary {\it uniformly}
within a certain range of values.
If the probability distribution for choosing $\alpha$ values is Gaussian
similar results are obtained, were the standard deviation $\sigma$ of the
distribution plays a similar role as the width $\Delta\alpha$ used
in Fig.~\ref{fig4}.

However, if we take values within a range, say $\alpha\in[0.1,0.9]$, and 
chose them according to a power-law distribution $P(\alpha)\sim 
\alpha^{-\beta}$ the fractal dimension changes significantly, as 
shown in Fig.~\ref{fig5}.
Even for small values of the exponent $\beta$, e.g.~$\beta=0.5$,
the fractal dimension decreases when compared to the value obtained
for the uniform distribution ($\beta=0$) and remains approximately
constant at $d_f\sim 1.43$. 

Notice that $\beta=0$ in Fig.~\ref{fig5} corresponds to $\Delta\alpha=0.8$
in Fig.~\ref{fig4}b. If the power-law distribution selects
values in a range with a different width, a similar decrease of the fractal
dimension is observed when comparing with the uniform distribution case.
Therefore, one can conclude that a reasonable choice for constructing
space-filling bearings with fractal dimension similar to the one observed
in fault gouges is by taking a random value of $\alpha$ uniformly
distributed in a certain range around $0.5$.

\section{The three-dimensional case}
\label{sec:3D}

As described above in Sec.~\ref{sec:prevalg}, a three-dimensional
version of fully random space-filling bearings is obtained in a similar
way as for discs with the single difference that the introduction of new 
spheres takes into account four nearest neighbors.
In this Section we address the case of three-dimensional space-filling
bearings as a more realistic approach to fault gouges, and study
how well the two-dimensional model approximates three-dimensional
systems of spheres.
\begin{figure}[b]
\begin{center}
\includegraphics*[width=8.5cm]{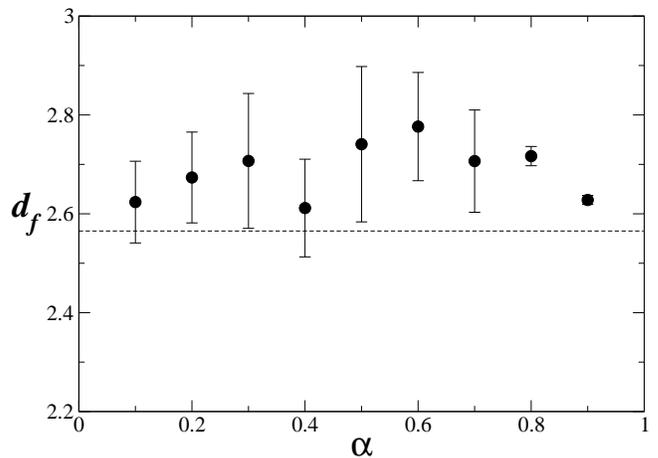}
\end{center}
\caption{\protect 
  The fractal dimension as a function of $\alpha$ for three-dimensional 
  random space-filling bearings when $\alpha$ is kept constant.
  A similar procedure as the one illustrated in Fig.~\ref{fig2} is
  used, with new spheres being introduced touching the {\it four}
  nearest neighbors (see text). The dashed line indicates one typical
  value $d_f\sim 2.58$ found in some real fault gouges~\cite{abe05}.}
\label{fig6}
\end{figure}
\begin{figure*}[t]
\begin{center}
\includegraphics*[width=15.0cm]{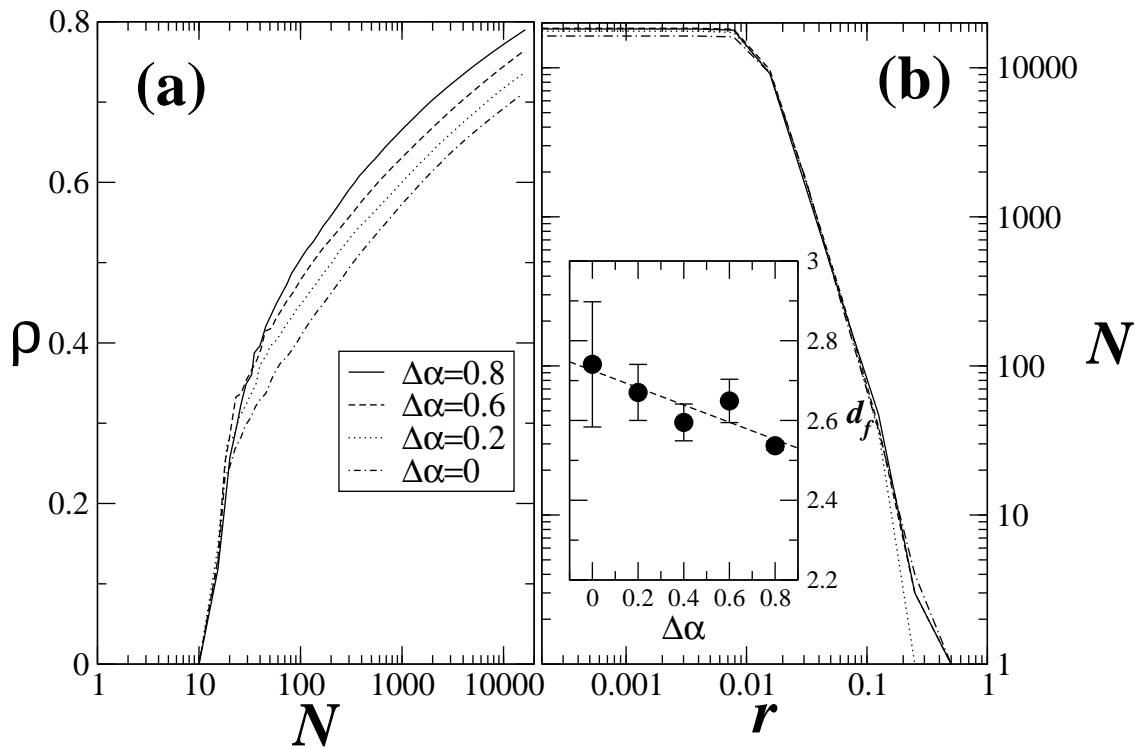}
\end{center}
\caption{\protect 
  Three-dimensional space-filling bearings for varying 
  $\alpha\in [0.5-\Delta\alpha/2,0.5+\Delta\alpha/2]$.
  {\bf (a)} The density $\rho$ as a function of the number $N$
  of spheres for $\Delta\alpha=0,0.2,0.6$ and $0.8$ and
  {\bf (b)} the corresponding size distribution $N(r)$ with the
  fractal dimension $d_f$ in inset. In all cases $N=10^4$.}
\label{fig7}
\end{figure*}

Recently~\cite{abe05}, it was found that grain fracture simulations 
produce a comminuted granular material similar to the one observed
in real fault gouges. 
From those simulations, it followed that comminution rate and survival 
of large grains is sensitive to applied normal stress, with a fractal 
dimension of the resultant grain size distributions in the range 
$d_f\in [2.3\pm 0.3, 2.9 \pm 0.5]$, that agrees with the observations 
of three-dimensional samples of real gouges where typically
$d_f\sim 2.58$.

In three dimensional space-filling bearings with a constant value
of $\alpha$ the fractal dimension lies above the observed values
in fault gouges. In Fig.~\ref{fig6} we plot typical values of $d_f$
as a function of $\alpha$. 
The fractal dimension of such bearings is typically larger than
$d_f\gtrsim 2.58$ (dashed line) with values within the range 
$d_f\in [2.60\pm 0.10,2.74\pm 0.15]$ and, similarly to the 
two-dimensional case, the maximum of $d_f$ is reached for 
$\alpha\sim 0.5$.

As summarized above in the introduction, it was recently 
found~\cite{sammis07} that fractal dimensions $\sim 2.6$ are 
observed for low-strain gouges.
In regions subject to larger shear strain the fractal dimension
is significantly larger, $\lesssim 3$.
Therefore, the particle size or mass dimensions were proposed 
as a way to distinguish between regions with different strain
strengths~\cite{sammis07}.
From Fig.~\ref{fig6}, one sees that a similar range of values 
for the fractal dimension is also found for space-filling bearings.
\begin{figure}[htb]
\begin{center}
\includegraphics*[width=8.5cm]{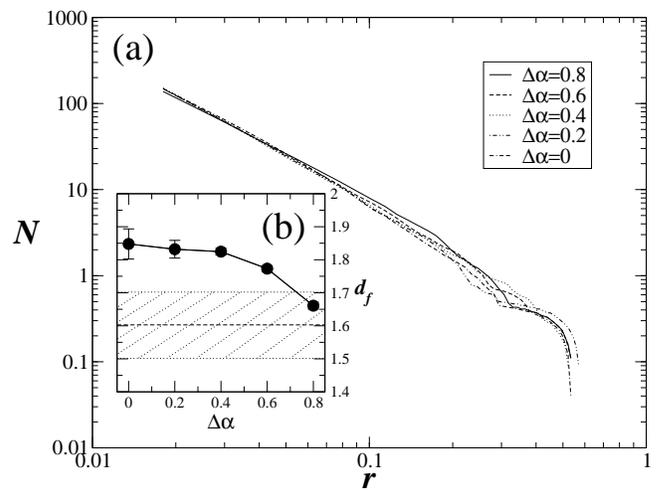}
\end{center}
\caption{\protect 
  {\bf (a)} The size distribution of two-dimensional cuts of the 
  three-dimensional space-filling bearings addressed in Fig.~\ref{fig7}
  and 
  {\bf (b)} the corresponding fractal dimension.}
\label{fig8}
\end{figure}

Furthermore, the explanation relating the fractal dimension of 
fault zones and their strain strength assumes that fragmentation
is controlled by nearest neighboring particle contact and
that a particle is most likely to split into smaller particles
with a particle of similar size, yielding a larger
fractal dimension $\sim 3$.
In the case of our construction procedure for space-filling
bearings this would correspond to the case of $\alpha\sim 0.5$.
Indeed, from Fig.~\ref{fig6} one observes that the maximum
of the fractal dimension is reached for such $\alpha$ values
yielding $d_f=2.74\pm 0.15$.

When varying $\alpha$ randomly in a range around $0.5$ and study
the dependence of the space-filling bearing on the width
$\Delta\alpha$ of the range $[0.5-\Delta\alpha/2, 0.5+\Delta\alpha/2]$.
In Fig.~\ref{fig7}a one sees that the density increases faster for larger
$\Delta\alpha$, similarly to what was shown in Fig.~\ref{fig4}a.
As for the fractal dimension, Fig.~\ref{fig7}b shows that it decreases 
slightly when compared with the case of constant $\alpha$ 
($\Delta\alpha=0$). Therefore, increasing the width $\Delta\alpha$
of the range of admissible values for $\alpha$ one is able to reduce
the fractal dimension of the bearing. 

Similarly to the situation of measures taken in fault gouges, the 
two-dimensional cross section of such three-dimensional space-filling
bearings should have a fractal dimension within the range of the
observed empirical values in real fault gouges 
($d_f=1.6\pm 0.1$~\cite{sammis87,sammis89}). By averaging several different
two dimensional cross sections of the 3D bearings 
we plot in Fig.~\ref{fig8}a the size distribution of a typical 
two-dimensional cross section for the different values of $\Delta\alpha$.
In Fig.~\ref{fig8}b we observe that only for very wide ranges of $\alpha$
values it is possible to obtain a fractal dimension similar to the one 
observed on fault zones.

\section{Discussion and conclusions}
\label{sec:conc}

In this work we studied the size-distribution of random space-filling 
bearings with large polydispersity, showing that it reproduces well
the size-distribution found in fault gouges.
Focusing on the dependence of the bearings fractal dimension on the 
spacing offset, we have shown that the fractal dimensions of such 
bearings sweep the low range of values observed in real faults.
Since recently it has been reported that the fractal dimension
varies in space along fault gouges\cite{sammis07}, our findings enables 
us to 
conjecture that the occurrence of seismic gaps, where earthquakes
are absent and therefore behave similarly to roller bearings,
may occur in regions where the fractal dimension lies in the low
range of admissible values, namely $d_f\in [2.5,2.75]$.

To compute an accurate value for the exponent characterizing 
random bearings, we introduced a general algorithm that allows 
$\alpha$ to vary randomly in a wide range of admissible values, typically 
$0< \alpha< 1$ and start the space-filling procedure from one 
unique large disc (or sphere), maximizing the range of admissible sizes
in the bearing.

With such model we
were able to show that bearings have a fractal dimension with values
within the range of values in real fault. 
Since it is known~\cite{sammis07} that along a specific fault gouge
the fractal dimension varies typically between $\sim 2$ and
$\sim 3$, our results support the hypothesis that seismic
gaps, occurring only in certain particular locations of the
fault, could be explained by this simple geometrical model.

Further, we also interpret the control parameter $\alpha$ for
the bearing property as a measure of the fragmentation strength,
and introduce simple criteria to improve the computational
efficiency of previous space-filling packing algorithms.

To improve further our findings we should also take the
effect of gravity into account. 
Moreover, concerning the space-filling bearings by themselves,
other questions raise, namely their contact network correlations,
which should help to understand the range of observed fractal
dimensions.
These and other points will be addressed elsewhere.

\section*{Acknowledgments}

The authors thank Bibhu Biswal for useful discussions.
This work was supported by {\it Deutsche Forschungsgemeinschaft},
under the project LI 1599/1-1.


\end{document}